\documentclass[a4paper,fleqn,usenatbib,useAMS]{mn2e}
\usepackage{graphicx}	
\usepackage[flushleft]{threeparttable}
\usepackage{amsmath}	
\usepackage{multicol}        
\usepackage{bm}		
\usepackage{pdflscape}	
\usepackage{color}
\usepackage{hyperref}
\usepackage{natbib}
\usepackage[T1]{fontenc}
\usepackage{ae,aecompl}
\usepackage{newtxtext,newtxmath}
 
\def\aap{AA}

\def\apjl{ApJL}

\def\mnras{MNRAS}
\def\apj{ApJ}

\def\aj{AJ}
\def\jcap{JCAP}

\def\prd{Phys.Rev.D}

\def\Ok{{\Omega_{k}}}

\title[$H_0$ and $r_s$ from low-redshift probes]{The Hubble-Lema{\^i}tre constant and sound horizon from low-redshift probes}

\author[ R.~Wojtak \& A.~Agnello]{Rados{\l}aw Wojtak$^{1}$\thanks{E-mail:
radek.wojtak@nbi.ku.dk}, Adriano Agnello$^{1}\thanks{E-mail:
adriano.agnello@nbi.ku.dk}$ 
\\
$^{1}$DARK, Niels Bohr Institute, University of Copenhagen, Lyngbyvej 2, 2100 Copenhagen, Denmark \\
}
\date{Last updated 2018 Nov. 13}
\pubyear{2018}
\hypersetup{draft}

\begin{document}
\label{firstpage}
\pagerange{\pageref{firstpage}--\pageref{lastpage}}
\maketitle

\begin{abstract}
We revisit the claimed tension, or lack thereof, of measured values of the Hubble-Lema\^{i}tre parameter $H_0$ from Cosmic Microwave Background (CMB) data and low-redshift indicators. Baryon Acoustic Oscillations (BAO) rely on the scale of the sound horizon at recombination $r_s$ to convert angular measurements into angular-diameter distances, so fixing $r_s$ from CMB measurements already constrains $H_0.$ If departures from concordance cosmology are to be constrained, truly independent measurements of $H_0$ are needed. We use the angular-diameter distances to three time-delay lenses from the H0LiCOW collaboration to calibrate the distance ladder, combine them with relative distances from Supernovae~\textsc{I}a and BAO, leaving $r_s$ completely free, and provide the inferred coefficients ($q_{0},j_{0},s_{0}$) in the polynomial expansion of $H(z).$ We obtain $H_{0}r_{s}=(9895\pm161)$~km$\;{\rm s}^{-1}$ and $H_0=(72\pm7)$~km$\;{\rm s}^{-1}\;{\rm Mpc}^{-1}$. Combined with $H_0$ from H0LiCOW, then $r_s=(137\pm4.5)$~Mpc is consistent with previous work and systematically lower than the CMB-inferred value. Our results are independent of the adopted cosmology, and removing Supernovae with $z<0.1$ has a negligible effect.

\end{abstract}


\begingroup
\let\clearpage\relax
\endgroup
\newpage

\section{Introduction}
\label{intro}
Using various observational data sets and methodologies to measure cosmological parameters is necessary in order to test our understanding of concordance cosmology and, ultimately, of fundamental physics. A stunning discovery of the (late-time) accelerated expansion of the Universe \citep{Riess1998,Per1999}, and the discovery and characterization of the Cosmic Microwave Background  \citep[CMB; e.g.][]{Penzias65,Smoot92} gave credence to the Big Bang $\Lambda$CDM cosmological model.

Despite its success as an overall effective description of currently available cosmological data, the model is yet to be challenged by various consistency tests, in particular tests of compatible behaviour of the model at low and high redshifts. In this context, comparing the measurements of the Hubble-Lema\^{i}tre parameter $H_{0}$ from low- and high-redshift probes of the expansion history has recently drawn particular attention, in large extent driven by progress in observations at extremely low and high redshifts.

CMB experiments cannot measure $H_0$ directly, but enable an inference once the other cosmological parameters are chosen or determined in a joint fit to CMB data. The inference, however, depends on the choice of cosmological model. On the other hand, distance indicators at lower redshift can be used to measure $H_0$ and hence constrain possible departures from concordance cosmology when compared with CMB predictions. Recent observations of Type~\textsc{I}a~Supernovae (SNe) \citep[calibrated with Cepheids,][]{Riess2018}  and time-delay lenses \citep{birrer18} yield 
measurements of $H_0$ that are systematically higher than the CMB predictions of a flat $\Lambda$CDM cosmological model constrained by the CMB data from the Planck satellite mission \citep{Planck2018}.

A recent claim \citep{mac18} has been made that $H_0$ agrees with the CMB inference once Baryon Acoustic Oscillations (BAO) and Supernovae are used, in a so-called \textit{inverse distance ladder} formalism. Before that, Barnal et al. (2016) and Aylor et al. (2018) had already used BAO distances, calibrated with Cepheids or $H_0$ from time-delay lenses, to infer the sound horizon at recombination $r_s.$ Using either $\Lambda$CDM or spline models for the expansion history $H(z),$ those authors obtained $r_s=(138\pm4)$~Mpc, with little dependence on the chosen cosmological models. The apparent discrepancy between those earlier results and those of \citet{mac18} is because in the former $r_s$ was a free parameter to be inferred, whereas in the latter it was anchored to CMB measurements.

Here, we revisit the calibration of BAO and Supernova distances and their use in the inverse distance ladder. We use angular-diameter distances to three time-delay lenses in order to calibrate the distance ladder, and the \textit{relative} behaviour of distances from BAO to infer $H_0$ and the sound horizon at recombination $r_s$ independently of CMB measurements. A distance ladder with SNe~\textsc{I}a and angular-diameter distances to two lenses, in the context of (departures from) $\Lambda$CDM, has been used by Jee et al.$^\dag$\footnote{$^\dag$ Jee \& Suyu, private comm.} (2018, subm. to Science).
Here we use the same cosmographic  parameterization of $H(z)$ and distances as in \citet{mac18}, for ease of comparison and to make our results independent of the adopted cosmological model.
We finally examine the reliability of different (relative) distance calibrators based on general relations between different distance measurements in cosmology. 

This Letter is organized as follows. The distance determinations are introduced in Section 2. Results on cosmological parameters are given in Section 3. We conclude in Section 4 and present a comparison of distance ratios from different cosmological probes, as well as forecasts on $r_s$ measurements from percent-level accuracy on low-redshift $H_0$ measurements. None of the following depends on the choice of a particular cosmological model, nor on the local calibration of the distance ladder from Cepheids. Throughout the following, we denote luminosity-, angular-diameter-, and comoving distances by $D_{\mathrm{lum}},$ $D_{M},$ $D_{\mathrm{ang}},$ respectively.

\section{Distance Indicators}
\label{methods}
Instead of working within a prescribed family of cosmological models, we follow \citet{mac18} and use fourth-order expansions in redshifts. In particular, the expansion history is parameterized as
\begin{eqnarray}
\frac{H(z)}{H_0} & \equiv & \frac{\dot{a}}{a\,H_0} = 1+(1+q_0)z+(j_0-q_0^{2})z^{2}/2\\
\nonumber & & +(3q_0^{3} +3q_{0}^{2} -4q_{0}j_{0}-3j_{0} -s_{0})z^{3}/6.
\end{eqnarray}
The luminosity distance $D_{\rm lum}(z)$ then follows
\begin{eqnarray}
D_{\rm lum}(z)\frac{H_{0}}{\mathrm{c}} & = & z+(1-q_{0})z^{2}/2\\
& & -(1+j_0-\Ok-q_{0}-3q_{0}^{2})z^{3}/6+d_{4}z^{4}/24,
\end{eqnarray}
with
\begin{eqnarray}
\nonumber d_{4} & = & 2+s_0+5j_{0}-2\Ok-2q_0\\
& & +10j_{0}q_{0}-6{\Ok}q_{0}-15q_{0}^{2}-15q_{0}^{3}\ .
\end{eqnarray}

Above, we have followed \citet{Vis2004} and incorporated curvature ($\Omega_{k}$) in the model \citep[see also][]{Wein2008}.

The other distances ($D_{\mathrm{ang}}$ and $D_{M}$) are calculated using the distance duality relations, i.e. $D_{\rm lum}=D_{\rm ang}(1+z)^{2}$ and  $D_{\rm lum}=D_{M}(1+z),$ which holds not only for general relativity, but also for a wide class of metric-based theories of gravity \citep{Eth1933,Bas2004,Sch1992}. Overall, our approach {to joint modelling of distance and expansion-history data} is therefore independent of the Einstein field equations.

\subsection{Supernovae~\textsc{I}a}
Type-\textsc{I}a SNe have long been used as standardizeable candles to obtain \emph{relative} luminosity distances over a wide redshift range. This enabled the discovery of the (late-time) accelerated expansion of the Universe. In contrast to the deceleration parameter, which is constrained primarily by relative distances at high redshifts, $H_{0}$ requires a distance calibration. The local \textit{distance ladder} calibration realized by the SH0ES project \citep{Rie2016}, using parallax distances to Cepheids in the Milky Way and in nearby SN hosts, yields $H_0=73.48\pm1.66$~km$\,$s$^{-1}\,$Mpc$^{-1}$ \citep{Riess2018}. The measurement is consistent with several alternative (though not equally precise) local distance calibrations based on observations of water maser emission in accretion disks \citep{kuo13,Rie2016} or the tip of the red giant branch \citep{Hatt2018}.

Here, we use the binned distance moduli from the Joint-Lightcurve-Analysis sample \citep[JLA,][]{Bet2014} which combines SDSS-II and Supernova Legacy Survey (SNLS) observations, taking into account a number of systematic effects related to photometry, light curve fitting and variation of the average absolute luminosity of 
Type-\textsc{I}a SNe with the properties of their host galaxies.

\subsection{Baryon Acoustic Oscillations}
 The sound horizon of acoustic oscillations at recombination imprints its characteristic scale $r_s$ in the large-scale structure. This can be measured at low redshifts through detection of peaks in the two-point correlation function of galaxies. The comoving BAO scale does not evolve with redshift, and so it can serve as a standard ruler to measure: distances, when the BAO is observed in clustering of galaxies in the sky; and the Hubble parameter, when the BAO scale is observed in clustering of galaxies in redshift space \citep[][]{Bas2010}. The conversion to absolute sizes requires prior knowledge of the BAO scale. An alternative approach, fully agnostic about constraints from the early Universe, would leave the BAO scale as a free parameter, thereby measuring only the product $H_{0}r_{s}.$
 
 When using BAO data to measure $H_{0},$ an independent distance calibration is required. A common approach relies on $r_{s}$ determined by physics of the early Universe, which can be constrained by precise observations of the CMB.  An alternative is to calibrate the BAO scale with low-redshift observations. Here, we choose to do so with angular-diameter distance measurements from time-delay lenses (spanning $0.3\lesssim z\lesssim 0.8$). 

We use the anisotropic BAO measurements by the Baryon Oscillations Spectroscopic Survey (BOSS). We adopt consensus pre-reconstruction best fit distances and $H(z)r_s$ in three redshift bins centered at $z=0.38$, $z=0.51$ and $z=0.61$, and the corresponding covariance matrix, from \citet{Ala2017}. We also include the isotropic BAO signal determined from a joint analysis of the 6dF Galaxy Survey \citep{Beu2011} and the SDSS Main Galaxy sample \citep{Ros2015}, in particular the volume-averaged distance $D_{V}=[\mathrm{c}zD_{M}(z)^{2}/H(z)]^{1/3}$ at redshift $z=0.122$ \citep{Car2018}. For the sake of completeness, we also consider constraints on $D_{V}$ at $z=0.44,\,0.6,\,0.73$ from WiggleZ \citep{Kaz2014}.

The BOSS measurement has the largest constraining power among all BAO observations used in our study. The consensus measurements obtained by \citet{Ala2017} include  systematic errors arising from the choice of fiducial mock galaxy catalogues and fitting methods.

\subsection{Time-delay lensing}
In strong lensing, the light from background sources is deflected by intervening mass to the extent that multiple images arise. If the source is variable, the light-paths through different images result in a measurable time-delay between the observed lightcurves \citep{Refsdal64}, $\Delta t=(1+z_{l})\frac{D_{l}D_{s}}{D_{ls}}\Delta\psi/\mathrm{c},$ which depends on the relative distances$^\dag$\footnote{Here $D_{l}=D_{\rm ang}(0,z_{l}),$ $D_{s}=D_{\rm ang}(0,z_{s}),$ $D_{ls}=D_{\rm ang}(z_{l},z_{s}).$} between observer, deflector and source, as well as the lensing configuration and gravitational potential of the deflector (summarized here in the Fermat potential difference $\Delta\psi$). This has been used by the H0LiCOW collaboration \citep{Suyu17} to infer $H_0$ from time-delay lenses with exquisite monitoring data from COSMOGRAIL\footnote{\texttt{https://cosmograil.epfl.ch}} \citep[][]{Tewes12}. The main source of uncertainty is from the adequacy of lens models, and the role of additional mass along the line of sight \citep{sch13}. When the velocity dispersion $\sigma$ of the deflector is measured, the ratio $\mathrm{c}^{3}\Delta t/\sigma^{2}\propto D_{l}$ makes these systems standard rulers independent of any cosmological model \citep{par09}, provided the degeneracies in lensing and dynamical models are suitably explored \citep{jee16,sha17}, regardless of weak lensing by additional line-of-sight perturbers.

\begin{table}
    \begin{tabular}{l|l|l|l}
    lens & $D_{\rm ang} [Mpc]$ & z & reference \\
    \hline
RXJ1131 & $813.33\pm113.9$ & 0.295 & \citet{jee15}\\
B1608 & $1485.7\pm208.0$ & 0.6304 & \citet{jee15}\\
J1206 & $1804.0\pm460.0$ & 0.745 & \citet{birrer18}\\
\end{tabular}
\caption{Angular-diameter distances to three lens galaxies. The distances were determined by means of combining gravitational time delay measurements with constraints on dynamical mass from stellar kinematics.}
\label{lenses}
\end{table}

Here, we use three angular-diameter distances, to set an absolute distance calibration which can then be extrapolated to $z \rightarrow 0$ using BAO and SNe~\textsc{I}a. Specifically, we adopt distances to two lenses as computed by \citet{jee15}, and to a third lens from \citet{birrer18}; see Table~\ref{lenses}.  We remark that these $D_{\rm ang.}$ measurements require working hypotheses on the velocity anisotropy tensor of stars in the deflectors. In particular, current measurements by H0LiCOW assume Osipkov-Merritt anisotropy profiles, and the final inference on distances changes if more general functional forms are chosen. Jee et al. (2018, subm.) have reassessed the distance determination to two lenses in this sample, using two-parameter anisotropy families
\begin{equation}
\beta(r)=\beta_{0}+\frac{(\beta_{1}-\beta_{0})r^{2}}{r^{2}+r_{a}^{2}}\ ,
\end{equation}
and fitted the full shape of the posterior on $D_{\rm ang}.$ Their quoted distances\footnote{Jee et al., subm. to Science; Jee \& Suyu, private comm.} are consistent with the already-published values (which we adopt here), but slightly smaller, and so result in slightly higher $H_0$ measurements. In order to reduce systematics from the mass-anisotropy degeneracy, spatially-resolved kinematic measurements are needed. This has been shown to produce percent-level accuracy with realistic error budgets \citep{sha17}.
\begin{figure}
\includegraphics[width=0.45\textwidth]{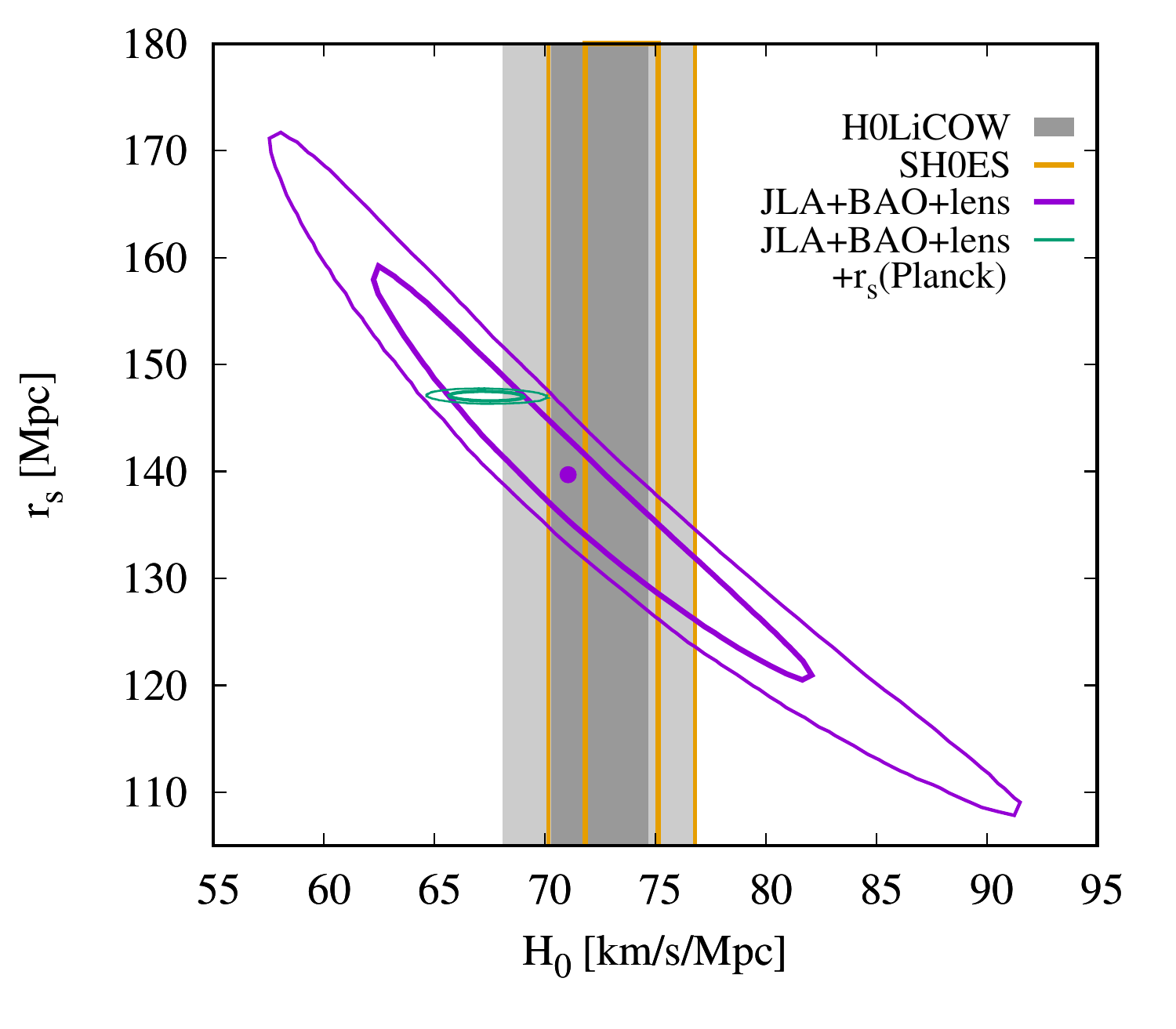}
\includegraphics[width=0.45\textwidth]{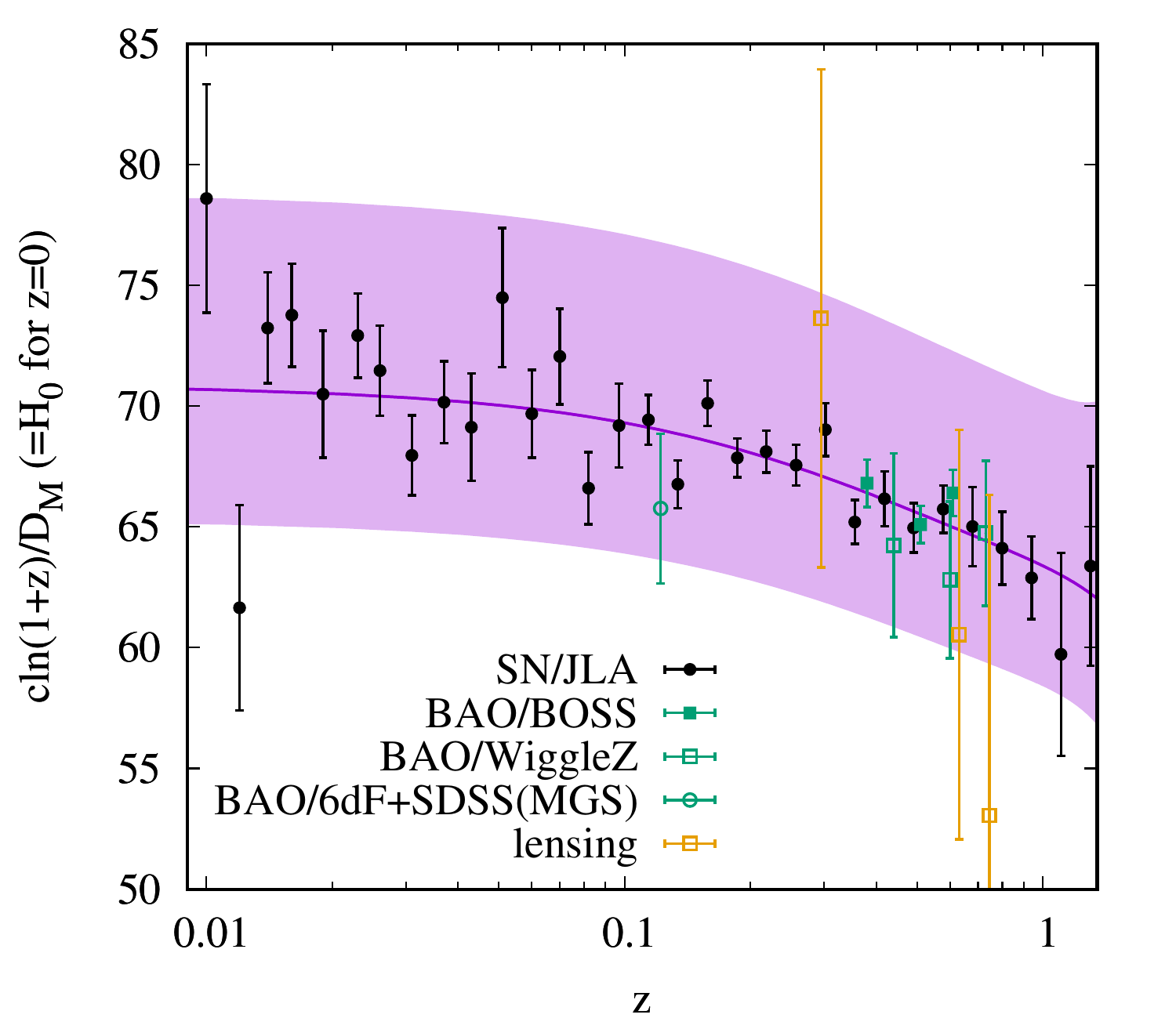}
 \caption{Cosmographic inference from the inverse distance ladder, calibrated on three angular-diameter distances from lensing and extrapolated via BAO and SNe~\textsc{I}a. \textit{The top panel} shows the combined inference on $H_0$ and $r_s,$ without priors on $r_s$ from the CMB. Both the CMB (green, $r_s\approx147$~Mpc) and low-redshift measurements H0LiCOW (grey bands) and SH0ES (orange lines) are compatible within the current uncertainty level. \textit{The bottom panel} shows a distance-related quantity which equals to $H_0$ at $z=0$. In this panel, the overall normalizations of BAO and SNe~\textsc{I}a distances are given by the best-fit model, which is determined in a joint fit including the lensing data.
}
\label{fig:h0rs}
\end{figure}
\subsection{Data analysis}
We explore cosmological parameter space by sampling the likelihood $\mathcal{L}$ defined through
\begin{eqnarray}
\ln \mathcal{L} & \propto & -\frac{\chi^{2}}{2} =  -\frac{1}{2}[\bm{\mu_{\rm SN}}-\bm{\hat{\mu}_{\rm SN}}(\bm{\theta})]^{T}\bm{C_{\rm SN}}^{-1}[\bm{\mu_{\rm SN}}-\bm{\hat{\mu}_{\rm SN}}(\bm{\theta})] \nonumber \\
 & & -\frac{1}{2}[\bm{D_{\rm BAO}}-\bm{\hat{D}_{\rm BAO}}(\bm{\theta})]^{T}\bm{C_{\rm BAO}}^{-1}[\bm{D_{\rm BAO}}-\bm{\hat{D}_{\rm BAO}}(\bm{\theta})] \nonumber \\
 & & -\frac{1}{2}[\bm{D_{\rm lens}}-\bm{\hat{D}_{\rm lens}}(\bm{\theta})]^{T}\bm{C_{\rm lens}}^{-1}[\bm{D_{\rm lens}}-\bm{\hat{D}_{\rm lens}}(\bm{\theta})]  \\
 \nonumber
\end{eqnarray}
where $\,\hat{}\,$ indicates a model observable, $\bm{\mu_{\rm SN}}$ is a vector of distance moduli from Type-\textsc{I}a SN data, $\bm{D_{\rm BAO}}$ is a vector including distances ($D_{\rm M}$ or $D_{\rm V}$) and Hubble parameters from BAO observations (for a fiducial value of $r_{\rm s}$), $\bm{D_{\rm lens}}$ is a vector of 
angular-diameter distances from the three gravitational lenses, {\bf C}$^{-1}$ is the inverse covariance matrix and $\bm{ \theta}=(H_{0},r_{\rm s},\Omega_{k},q_{0},j_{0},s_{0},M)$ is a vector of model parameters. Parameter $M$ is a free normalization of the SN Hubble diagram and it combines the absolute magnitude \citep[relative to the fiducial absolute magnitude from][]{Bet2014} and the logarithm of $H_{0}$. Supernova data alone do not constrain $r_{s}$ and $H_{0}$. Their role in a joint fit is to propagate constraints from BAO and lensing data between different redshifts, and eventually extrapolate them to $z=0$. BAO measurements from the three surveys (SDSS/6dF, BOSS and WiggleZ) are based on independent galaxy samples, and thus we assume no correlations between these three data sets. All presented credible intervals and contours were computed using the standard Monte Carlo Markov Chain technique.

\section{Results}
\label{results}
\begin{table*}
    \centering
    \begin{tabular}{l|c|c|c|c}
    data & $H_{0}$~[km s$^{-1}$ Mpc$^{-1}$] & $r_{s}$~[Mpc] & $H_{0}r_{s}$~[km s$^{-1}$] & $\chi^{2}_{\rm red}$\\
      \hline
 3 lenses + JLA+BOSS (pre-reconstruction) &
$72.3\pm6.9$ & $139.2\pm 13.3$ & $9975\pm 177$ & 1.14\\
  3 lenses + JLA+BOSS+6dF(+MGS)+WiggleZ & $72.0\pm 6.7$  & $138.6\pm 13.0$ & $9895\pm161$  & 1.06 \\
  3 lenses + JLA+BOSS ($z>0.1$) &$71.7\pm 6.8$ & $139.2\pm 13.1$ & $9894\pm211$ & 1.01 \\
3 lenses + JLA+BOSS (post-reconstruction) & $72.2\pm 6.8$  & $139.7\pm 13.0$ & $9995\pm165$ & 1.09\\
    \end{tabular}
    \caption{Inferred $H_0,$ $r_s$ and $H_{0}r_{s}$ for different combinations of data sets and goodness of fit quantified by the reduced $\chi^{2}$. Differences are at sub-percent level. These measurements are based on angular-diameter distances to three time-delay lenses, and do \textit{not} include direct $H_0$ measurements from H0LiCOW time-delay distances.}
    \label{table1}
\end{table*}
\begin{figure*}
    \centering
    \includegraphics[width=0.95\textwidth]{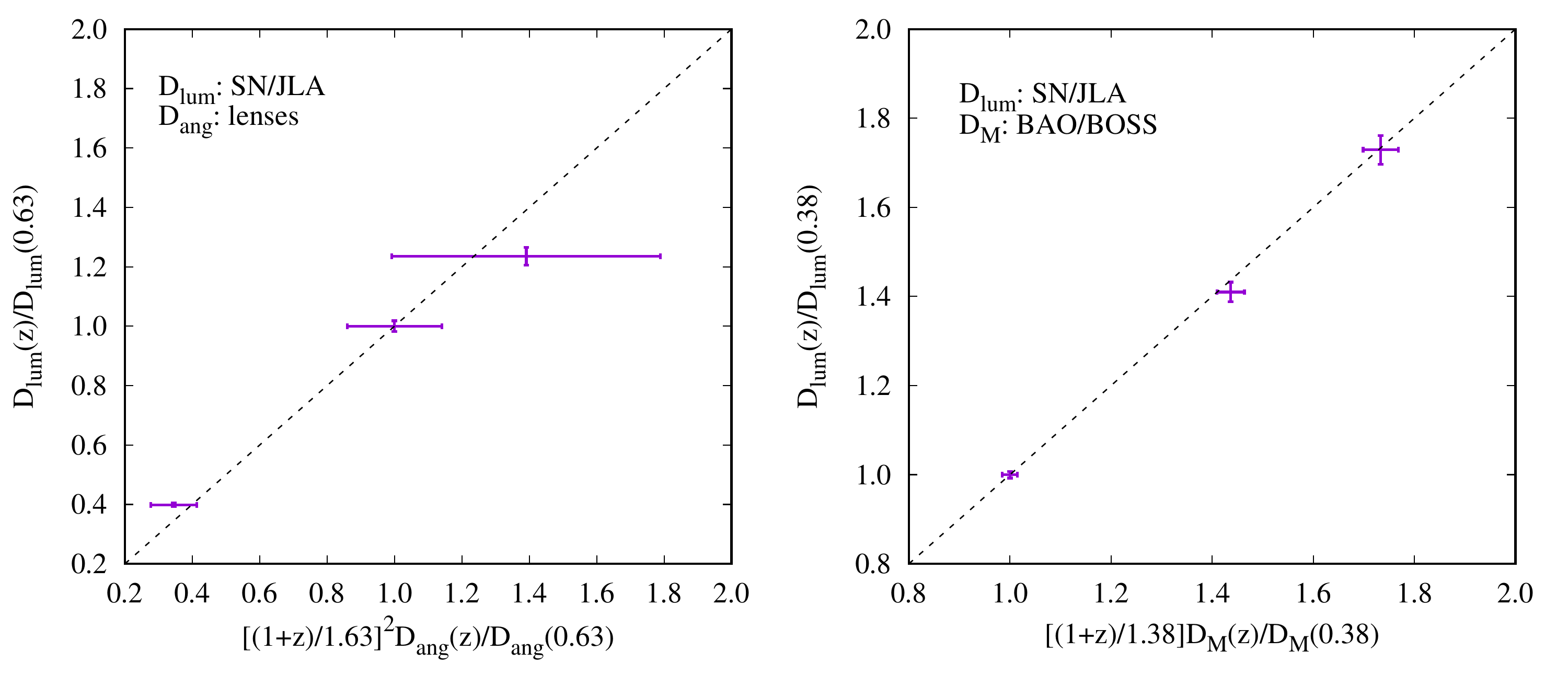}
    \caption{Consistency between relative luminosity distances from Type-\textsc{I}a SNe, angular diameter distances from lensing observations (\textit{left}) and comoving distances from BAO observations. The dashed lines indicate equal values of the quantities on both axes.}
    \label{fig:my_label}
\end{figure*}

As a sanity check, we follow \citet{mac18} and start by fitting the BAO measurements, with $r_s=(147.05\pm0.30)$~Mpc from the CMB \citep{Planck2018}, and extrapolate via SNe~\textsc{I}a. The resulting parameters are: $H_0=(67.3\pm1.1)$~km$\,$s$^{-1}\,$Mpc$^{-1}$, $\Omega_k=0.24\pm0.32,$ $q_0=-0.47\pm0.14,$ $j_0=1.0\pm1.4,$ $s_0=1.2\pm3.2$. Quite unsurprisingly, these are broadly consistent with those displayed by \citet{mac18}, in particular $H_0=(67.1\pm1.3)$~km$\,$s$^{-1}\,$Mpc$^{-1}$.

Then, we leave $r_s$ completely free, and use the three published angular-diameter distances to H0LiCOW lenses as distance calibrators. The resulting best fit parameters are 
$\Omega_k=0.26\pm0.33,$ $q_0=-0.47\pm0.15,$ $j_0=1.01\pm1.41,$ $s_0=1.35\pm3.3$. The resulting value of $H_0=(72\pm7)$~km$\,$s$^{-1}\,$Mpc$^{-1}$ is higher than that based on CMB priors on $r_s$ (see Figure~1), but it may change once the updated distance measurements by Jee et al.$^\dag$ (2018, subm.) are available.

We can robustly constrain the combination $r_{s}H_{0}=(9895\pm161)\,\mathrm{km}\,\mathrm{s}^{-1}.$ This is slightly lower than, but still compatible with, the values obtained by Aylor et al. (2018). The lower value is partly caused by the 6dFGS measurement. Adding WiggleZ data does not affect the results appreciably. Table~\ref{table1} lists best fit $r_{s}$ and $H_{0}$ inferred from different combinations of data sets used in our study. Switching between pre- and post-reconstruction measurements of BAO from the BOSS data \cite[see more details in][]{Ala2017} has negligible impact on our results.

The distance calibration from lensing relies only on three angular-diameter distances. From the H0LiCOW collaboration, the time-delay distances $D_{\Delta t}=(1+z_{l})D_{l}D_{s}/D_{ls}\propto H_{0}^{-1}$ are also measured independently, yielding a tighter measurement of $H_0$ from lensing. With the current $H_0 = 72.5^{+2.1}_{-2.3}$~km$\,$s$^{-1}\,$Mpc$^{-1}$ from H0LiCOW \citep{birrer18}, the inferred sound horizon becomes $r_{s}=(137.0\pm4.5)$~Mpc.

Not by chance, this measurement is very close to the ones by Barnal et al. (2016) and Aylor et al. (2018). This holds despite the different parameterization of $H(z)$ and the use of a different choice of low-redshift distance calibration. Our data compilation only marginally favours non-flat models, and curvature has little-to-no impact on the best fit sound horizon and $H_{0}.$

Since SNe~\textsc{I}a extend down to very low redshift, concerns may be raised on their redshift accuracy from LSR corrections, and in general on whether the CMB reference frame or local-Universe reference frame should be used. With low-redshift velocities of $\approx1000$~km$\;{\rm s}^{-1}$, this may result in systematics $\delta z\approx0.3\%$ at $z=0.1$. Cosmological inference including low-z data may also be affected by cosmic variance implying $\delta H_{0}/H_{0}$ ranging from $0.3\%$ at $z<0.1$ to $0.9\%$ at $z<0.05$ \citep{Woj2014}. In order to check the robustness of our results to these two possible systematics, we repeat the analysis excluding $z<0.1$ SNe from the distance-ladder extrapolation. The credibility contours are just slightly wider, and none of the above results changes appreciably.

The error budget in our measurement of $H_{0}$ and $r_{\rm s}$ is clearly dominated by the uncertainties in the three angular-diameter distances from lensing observations. We estimate that the uncertainties in both BAO and SN data contribute to only 14 and 12 per cent of the final errors in $H_{0}$ and $r_{\rm s}$. The main source of uncertainties in the current estimates of distances to gravitational lenses stems from a limited precision of the velocity dispersion measurements and systematics from the mass-anisotropy degeneracy. Both effects, with a dominant contribution from the former, account for about 80 per cent of the total errors quoted in Table~\ref{lenses}\citep[see e.g.][]{jee15}.

\section{Discussion}
\label{discuss}
The Hubble-Lema\^{i}tre parameter $H_0$ and the sound horizon scale $r_s$ link the late-Universe and early-Universe epochs, so a cross-comparison of $H_0$ from CMB and $r_s$ from low-redshift indicators can shed light on new physics beyond the Standard Model, if uncertainties are well understood. Here, we have used inverse-distance-ladder extrapolations, using a fourth-order expansion in redshift, with absolute distance calibrations from time-delay lenses and relative distance calibrations from SNe and BAO. The best-fitting coefficients ($q_0,j_0,s_0,\Omega_k$) in the expansion and their credibility ranges are given in Section~3.

\subsection{The Sound Horizon from late-time probes}
From a model with uniform prior on $r_s$ and a polynomial expansion on $H(z)$ and distances, we obtain $H_{0}r_{s}=(9895\pm161)$~km$\;{\rm s}^{-1}$. This is independent of the absolute distance calibration, as it requires only the relative scaling of distances with redshift. With current data, if angular-diameter distances to three lenses are used in the inverse distance ladder, $H_0=(72\pm7)$~km$\,$s$^{-1}\,$Mpc$^{-1}$.

Besides the absolute distance calibration, given by $D_{\rm ang.}$ to three lenses, the H0LiCOW collaboration also provided time-delay distances to four lenses, yielding $H_0=72.5^{+2.1}_{-2.3}$~km$\,$s$^{-1}\,$Mpc$^{-1}$. Combined with our inference, we obtain $r_{s}=(137.0\pm4.5)$~Mpc. To ascertain a possible `tension' between low-redshift and high-redshift $r_s$ measurements, percent-level measurements of $H_0$ from low-redshift probes are needed.

Our findings are consistent with those of \citet{ber16}, who fitted (relative) luminosity distances from SNe~\textsc{I}a and a distance ladder calibration with Cepheids using a model-independent reconstruction of the expansion history and found $r_{s}=(136.8\pm 4)$~Mpc for flat space and $r_{s}=(133.0\pm 4.7)$~Mpc for fits with a free curvature. Adopting a similar methodology, but using updated BAO measurements, \citet{Ver2017} found $r_{s}=(140.8\pm4.9)$~Mpc with a free curvature parameter, fully consistent with our results. Interestingly, when using cosmic chronometers \citep{Jim2002} to set $H_0,$ the inferred sound-horizon scale increases appreciably, i.e. $r_{s}=(150.0\pm 4.7)$~Mpc.

The `tension' between low-redshift and high-redshift measurements of $r_s$ is lower in our case than in the analysis by \citet{ber16}. However, both low-redshift determinations are consistent with one another. Future improvements of distance calibrations based on the standard rulers from lensing observations have the potential to verify whether systematics from Cepheids can play a major role in any claimed tension, and if the apparent discrepancy arises from incomplete knowledge of fundamental physics or more general systematics in low-redshift distance measurements and extrapolation. Potential solutions to alleviate the tension, based on ad hoc extensions of the standard model, involve additional relativistic species \citep{ber16,Aylor18,pou18}, dark energy with equation of state $w<-1$ or dynamical dark energy \citep{val2016,Zhao2017,Luk2018}. Discrepant distance calibrations may also point to inadequacy of the standard Friedmann-Lema\^{i}tre-Robertson-Walker paradigm \citep[see e.g.][]{Woj2017}.

\subsection{Reliability of Low-Redshift Probes}
All of the above measurements rely on fundamental relations between cosmological distances. In particular, in utter generality $D_{\mathrm{lum}}=(1+z)D_{M},$ and $D_{M}=(1+z)D_{\mathrm{ang}}.$ Therefore, the ratios of different distance measurements corresponding to the same redshift pairs should scale in a known way. A low-redshift version of this test ($0.1\lesssim z\lesssim 0.3$) has been made on distances to SNe measured from different methods \citep{epm18}.

Figure~2 shows distance ratios for the BAO and lensing indicators used in this paper, rescaled with the appropriate $(1+z)$ factors, compared to luminosity-distance ratios from SNe~\textsc{I}a. From the current number of indicators, everything is consistent within the uncertainties. In future samples, amounting e.g. to $\approx 10-40$ time-delay lenses (Shajib et al. 2018, Jee et al. 2017), systems with high systematic uncertainties will be identifiable as outliers in these distance-ratio relations.

We note that the volume-averaged scale $D_{V}$ from BAO cannot be compared directly to comoving distances $D_{M},$ since $D_{M}(z)=\sqrt{D_{V}(z)^{3}H(z)/\mathrm{c}z},$ which would make this test dependent on the adopted expansion history $H(z)$. In particular, for the $D_M$ measurement from 6dFGS to be consistent with other distance scalings, $H_0=(60\pm6)$~km$\,$s$^{-1}\,$Mpc$^{-1}$ would be needed. This, besides their very narrow prior on $r_s$ from CMB, may have contributed to lead \citet{mac18} to infer a lower value of $H_0.$

\subsection{Masses and General Relativity}
The measurement of distances from a combination of lensing and kinematic data relies on the hypothesis that lensing masses and dynamical masses are equal. This holds true in General Relativity, but many non-standard models of gravity envisioned to explain the accelerated expansion (such as $f(R)$ gravity, see e.g. \citealp{Hu2007}) predict scale-dependent ratios of the two masses. This has been the subject of observational tests \citep[see e.g.][]{piz16,Col2018}. With the current accuracy of dynamical mass measurements, the uncertainties are $\approx20\%$ on individual systems.

\section*{Acknowledgments} 
We thank S.~H. Suyu, I. Jee, F. Courbin and T. Collett for interesting discussions before and/or during the preparation of this Letter. We also 
thank Nikki Arendse for invaluable and insightful comments. The authors were supported by a grant from VILLUM FONDEN (project number 16599). This project is partially funded by the Danish council for independent research under the project ``Fundamentals of Dark Matter Structures'', DFF--6108-00470.




\bsp	
\label{lastpage}
\end{document}